\documentclass[a4paper]{jpconf}
\usepackage{graphicx}

\def\la{\lower.5ex\hbox{$\; \buildrel < \over \sim \;$}}
\def\ga{\lower.5ex\hbox{$\; \buildrel > \over \sim \;$}}

\begin{document}
\title{The natural science of cosmology\\{\normalsize International Conference on Gravitation and Cosmology, Goa, December 2011}}

\author{P J E Peebles}

\address{Joseph Henry Laboratories, Princeton University, Princeton NJ 08544, USA}

\ead{PJEP@Princeton.edu}

\begin{abstract}
The network of cosmological tests is tight enough now to show that the relativistic  Big Bang cosmology is a good approximation to what happened as the universe expanded and cooled through light element production and evolved to the present. I explain why I reach this conclusion, comment on the varieties of philosophies informing searches for a still better cosmology, and offer an example for further study, the curious tendency of some classes of galaxies to behave as island universes.
\end{abstract}

\section{Introduction}

Cosmology aims to draw large conclusions from exceedingly limited evidence, but I have to conclude, to my surprise, that the considerations reviewed in Section~2 make a convincing case that the relativistic Friedman-Lema\^\i tre expanding world model is a good approximation. Cosmology is a work in progress, of course, and many lines of research are aimed at establishing a still better theory. I offer in Section 3 thoughts on directions this research has taken and likely will take. One direction is the search for anomalies where the reconciliation of theory and observation seems particularly difficult. A good anomaly will teach us something, either a better  understanding of the standard model or clues to an improved one. I offer in Section~4  an example, the apparent behavior of pure disk and elliptical galaxies as island universes.

\section{The cosmological tests}

The standard $\Lambda$CDM cosmology assumes general relativity theory. This is an extrapolation of some 14 orders of magnitude in length scale from the precision tests on the scales of the Solar System and smaller. It assumes that 95\% of the present mass of the universe is in two hypothetical forms, dark matter and dark energy. And it assumes initial conditions that agree with simple versions of the inflation picture of the very early universe, which is encouraging, but inflation is an idea, not a theory. Substantiating this considerable variety of bold assumptions is a demanding challenge. But the considerations reviewed here lead me to conclude that the challenge has been met, largely.

Detecting the relativistic curvature of the redshift-magnitude relation \cite{Riess,Perlmutter} is a rightly celebrated contribution to the cosmological tests. But one gets some feeling for the challenge of substantiating cosmology by imagining the situation in an alternative universe in which the only significant cosmological test is the redshift-magnitude relation. That would include the demonstration of a close to linear relation between redshift and distance at low redshift, which is in line with Einstein's argument for homogeneity. I expect the demonstration of the departure from linearity at high redshift by the SNeIa measurements would split the  cosmology community in this alternative universe into three camps. Those who already favored the Steady State cosmology on philosophical grounds would  point out that  the curvature is in line with the Steady State prediction, and would rightly claim strong support for this cosmology. Among those who had a philosophical preference for the relativistic Friedman-Lema\^\i tre  cosmology one camp would add Einstein's cosmological constant $\Lambda$ to their cosmology. The other camp would point out that the indicated value of $\Lambda$ is absurdly different from any reasonable estimate from well-established quantum physics. They would point out that in the Friedman-Lema\^\i tre  cosmology the universe at redshift $z>1$ was different from now, and supernovae of type 1a then therefore had to have been at least somewhat different from present-day ones. They would acknowledge that carefully explored tests have given no indication of evolution of the properties of SNeIa, but still insist that supernovae cannot be examined in close detail, and there is no way to exclude the possibility that some subtle evolution has escaped the tests. Each of these three camps has a reasonable case. 

\begin{figure}[h]
\begin{center}
\includegraphics[width=3.5in]{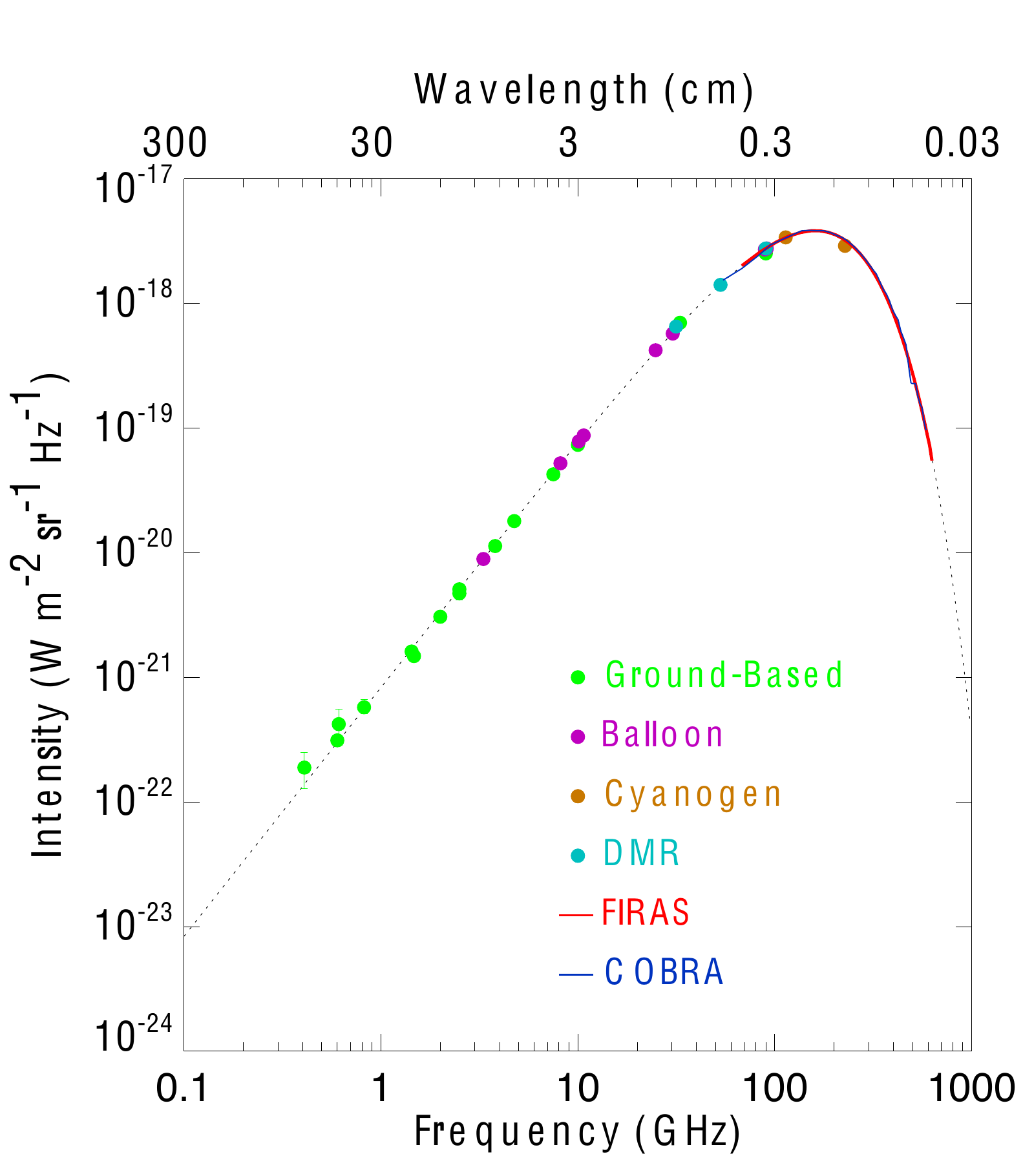}
\caption{Measurements of the CMB spectrum, compiled by Alan Kogut. The dotted curve is a thermal Planck spectrum.\label{Fig:CMB}}
\end{center}
\end{figure}

The point is that establishing cosmology requires a network of tests that check for systematic errors both in the measurements and in the theory used to interpret the measurements. For example, suppose cosmologists in the alternate universe had, in addition to the redshift-magnitude relation, measurements of the  cosmic microwave background radiation (the CMB), which show that the CMB is isotropic to better than a part in $10^5$ and that the spectrum is very close to thermal, as shown in Figure~\ref{Fig:CMB} (compiled and plotted by Al Kogut \cite{Kogut}, NASA GSFC). The only reasonable way to reconcile this measured thermal spectrum with the classical Steady State cosmology is to postulate that the universe is optically thick to the absorption and emission of microwave radiation at very small redshifts, for otherwise the CMB spectrum would be an unacceptable mix of temperatures set by the distribution of redshifts summed out past unit optical depth. Since people in the alternate universe could  could measure the CMB they would have detected discrete sources at CMB wavelengths. If they had adequate optical telescopes they would see that some are distant radio galaxies, meaning the universe is optically thin. This  serious challenge to the classical Steady State cosmology is an example of a contribution to a network of tests which, if dense enough, may establish cosmology. 

Returning to our universe, let us note that no natural science is complete, and that that certainly includes the established relativistic cosmology. For example, we have few constraints on what happened in the very early universe. One can save the Steady State philosophy by postulating that our universe behaves in a Quasi-Steady State manner \cite{Narlikar}, where the continual creation of the classical version is replaced by periodic bursts of creation at densities large enough to be capable of thermalizing the radiation. If the bursts of creation built up an appropriate distribution of matter and radiation at densities greater than that of the Big Bang at light element nucleosynthesis, and were followed by the expansion described by the relativistic model, then QSS would pass all cosmological tests and merit a place with variants of inflation in models for the very early universe that are viable but need work. 

Because the value of Einstein's $\Lambda$ in the standard Big Bang cosmology seems so unlikely within 
quantum physics it is natural and important to consider whether the evidence for detection of $\Lambda$ results from systematic error in the application of the relativistic cosmology.  An example is the proposal that the strongly inhomogeneous distribution of mass on scales less than a few tens of megaparsecs may significantly affect the general rate of expansion of the universe \cite{Kolb}. I believe Siegel and Fry \cite{SiegelFry} first answered this question. One may start with the observation that deflection angles, $\delta\theta$, in gravitational lensing are small, $\delta\theta\sim G\delta M/rc^2\la 10^{-5}$ around a mass concentration $\delta M$ of radius $r$. This angular deflection is a measure of the departure from the homogeneous Friedman-Lema\^\i tre geometry, $|\delta g_{\mu\nu}|\la 10^{-5}g_{\mu\nu}$. (Black holes are exceptions but irrelevant because at any distance of interest for this discussion they behave as ordinary particles.) One is invited to compute the spacetime geometry to second order in $\delta g_{\mu\nu}$ sourced by the mass autocorrelation function, which is reliably measured on the scales of significantly nonlinear mass clustering. The result (which requires a modest extension of the calculation in \cite{SiegelFry}) is that nonlinear mass fluctuations modeled as a gas of particles with mass $m$ and mean number density $\bar n$ appear in the Friedman equations as an effective contribution to the mean mass density, $\rho_{\rm eff} = m\bar n (1 + K + W)$, and to the pressure,  $p_{\rm eff} = m \bar n (2K + W)/3$, where $K= \langle v^2\rangle/2$ is the kinetic energy of peculiar motion per unit mass and $W = -{1\over 2}Gm\bar n a^2\int\,d^3x\langle\delta(\bf r)\delta(\bf r+ \bf x)\rangle /x$ is the gravitational energy per unit mass belonging to the departure from homogeneity. The effective pressure $p_{\rm eff}$ can be negative, as for $\Lambda$, but it is tiny, five orders of magnitude smaller than the standard estimate of $\Lambda$. This is not a way out of $\Lambda$ or dark energy.

Another proposed way out supposes that $\Lambda$ is small or zero and that the mass density and expansion rate vary with distance from us in such a way as to make the angular size distance vary with redshift in the same way as in a homogeneous universe with the standard value of $\Lambda$. We would have to be very near the center of an accurately spherically symmetric  universe, for otherwise the Sachs-Wolfe effect would produce an unacceptably large CMB anisotropy. Our favored position is arguably unlikely, but more more important is that this model is challenged by other cosmological tests. Intergalactic plasma scatters CMB photons, meaning the spectrum in Figure~\ref{Fig:CMB} is a mix of spectra originating at a range of distances. If the temperature at decoupling were not very close to homogeneous the CMB spectrum would be a mix of thermal spectra at different temperatures, contrary to the measurements. In the standard cosmology acoustic waves in the baryons and radiation prior to decoupling produced waves in the CMB anisotropy power spectrum and in the power spectrum of the large-scale galaxy distribution. Analyses \cite{Komatsu, Percival} pointing to consistency of the measurements of the CMB and galaxy distributions depend on, and test, many elements of the  cosmology, but my point  is that measurements of the angular distribution of the radiation probe what  happened near the edge of the observable universe and measurements of the spatial distribution of the galaxies probe what happened much closer to us. The consistency of these measurements within the homogeneous $\Lambda$CDM cosmology argues that conditions actually were quite similar across the sky at these two quite different ranges of distance, or else that there was yet another conspiracy. 

\begin{figure}[h]
\begin{center}
\includegraphics[width=14cm]{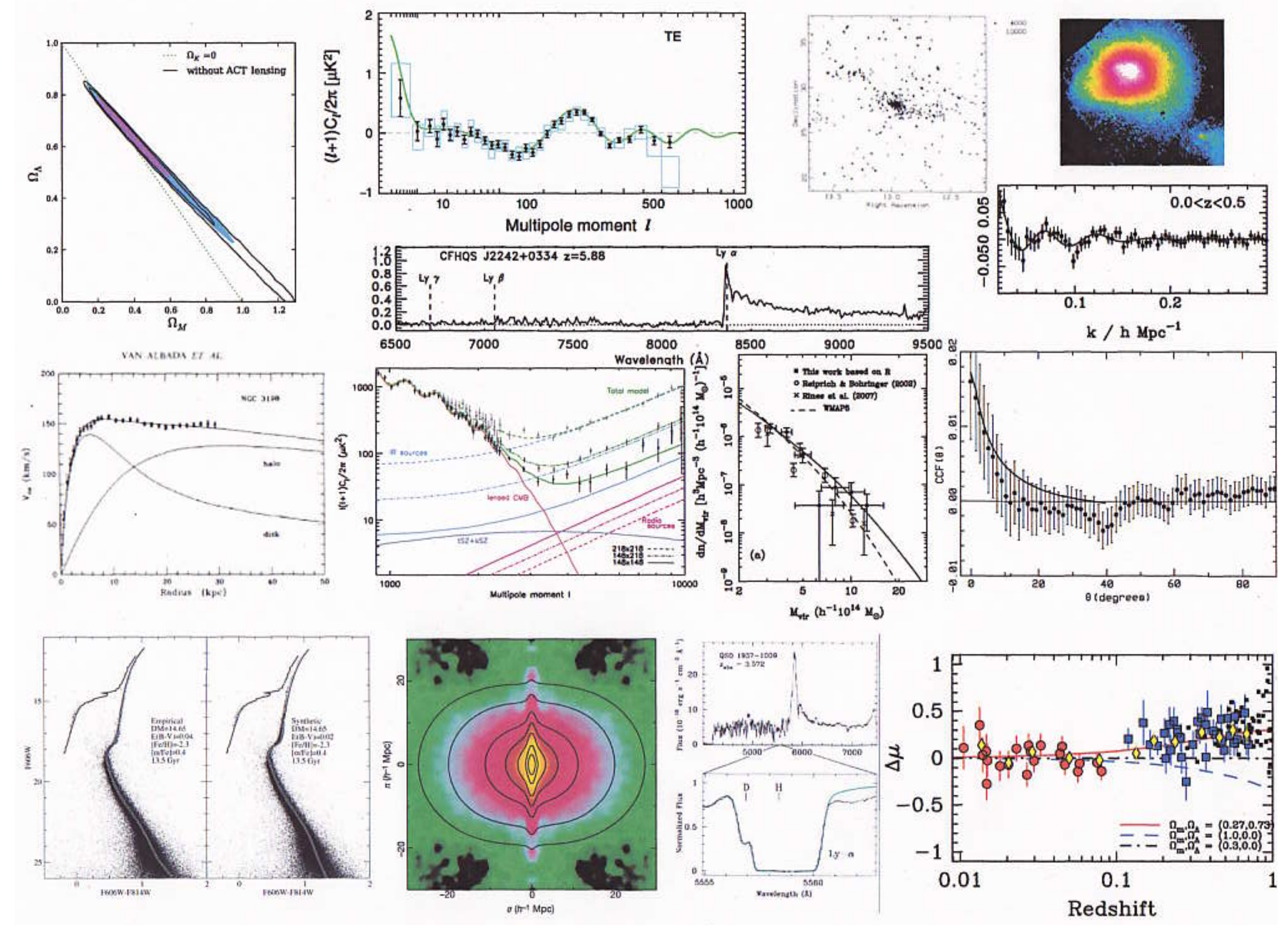}
\caption{Figures taken from papers on cosmological tests, for the purpose of illustrating the broad variety of phenomena and means of measurement and analysis in the network of tests.\label{Fig:tests}}
\end{center}
\end{figure}

This discussion could continue at length, but I instead refer to Figure~\ref{Fig:tests} for a schematic illustration of the broad variety of phenomena and means of measurement employed in the cosmological tests. The consistency of results from these many and varied approaches convincingly argues against the possibility that we have been seriously misled by systematic errors in the observations or errors in the theory by which the observations are interpreted. 

I hope it is understood that I am not arguing that cosmology after baryogenesis  is complete: that is contrary to the experience that natural sciences advance by successive approximations. But we can be sure that if there is a better cosmology it will predict a universe that looks much like $\Lambda$CDM, because the tests look at the universe from many sides now and find that it looks much like $\Lambda$CDM.

\section{The future}

We get some sense of the future of this subject by recalling how we got to where we are now. I see several paths common to cosmology and other parts of natural science. 

\noindent 1. {\it Serendipitous discovery of suggestive phenomena.} In relativity physics one might think of the apparently curious behavior of the velocity of light. In cosmology one thinks of the linear redshift-distance relation, which by 1936 Hubble and Humason \cite{Hubble} had tested out to redshift $z=0.1$. This relation was the centerpiece of most discussions of cosmology until the CMB, in 1965. Serendipity strikes seldom, but let us remember that it happens. 

\noindent 2. {\it The philosophical appeal of ideas.} Einstein followed this path to wonderfully good effect in his discovery of general relativity theory and his proposal that a philosophically satisfactory universe is homogeneous in the large-scale average. He had no justification for homogeneity from phenomenology, and only a few hints to GR --- electromagnetism and the E\"otv\"os experiment. It is striking that Nature on occasion agrees with our ideas of elegance. A contrary example is the fate of the classical Steady State cosmology, though the idea arguably reappears in eternal inflation. I  include as modern contenders cosmic strings, superstring cosmology, the multiverse and the anthropic principle. 

\noindent 3. {\it Mathematical incompleteness.} We take it as obvious that Nature abhors the singularities in general relativity. The community favorite remedy for the singularity in $\Lambda$CDM at infinite redshift  is a variant of  inflation, which needs work. 

\noindent 4. {\it Testing ideas.} The central example in cosmology is the program of cosmological  tests that has succeed so remarkably well, after some renegotiation of ideas about $\Lambda$. An influential idea driving research now is the possibility of a signature of inflation in the polarization of the CMB. A  detection  would have a wonderfully strong effect on the status of inflation. Laboratory searches may detect dark matter particles that behave much like the hypothetical cold dark matter in $\Lambda$CDM. If this detected matter along with massive neutrinos saturates the mass in matter in the dark sector this will close a line of research as far as cosmology is concerned. But resolution of an issue often raises new ones. Maybe some component of the dark matter will prove to do something interesting, leading us to new lines of research. The fascinating enigma of Einstein's $\Lambda$ motivates the study of  modifications of the physics of gravity and the  dark sector, and the research programs aimed at detecting or bounding evolution of $\Lambda$. Testing variation of $\Lambda$ is good science, though I must say I am uneasy about such heavy expenditure of resources on a  shot in the dark. Let us support it, but also support wider searches for cosmological tests.

\noindent 5. {\it Anomalies.} Two celebrated examples are in Lord Kelvin's essay \cite{Kelvin},  {\it Nineteenth Century Clouds over the Dynamical Theory of Heat and Light}. One cloud is exemplified by the curious behavior of mercury vapor. A spark causes the vapor to radiate at sharply defined frequencies  seemingly  scattered at random across the visible spectrum. The vapor must contain resonators that radiate at definite  frequencies when excited. The anomaly was that these radiators ought to be excited when the vapor is heated, yet the heat capacity of mercury vapor is quite close to that of an ideal gas of structureless particles. Lord Kelvin knew physics is well founded: he owed his fortune and peerage to his great contributions to physics, to say nothing of his income from physics-based patents for transatlantic telegraph cables among other things. My impression of Kelvin's position, put in modern jargon, is that  physics is solid and the rates for thermal excitation of internal energies of gasses must be  slow enough  that heat capacities were measured in systems that had not relaxed to statistical equilibrium. At the time Planck was struggling with another heat capacity problem, that of thermal radiation, another precursor of the perfect storm of new physics to come. Kelvin acknowledged that the other cloud in his essay, the curious behavior of the velocity of light, is ``very dense.'' He mentioned with approval ``the brilliant suggestion made independently by FitzGerald and by Lorentz of Leiden that the motion of ether through matter may slightly alter its linear dimensions.'' That was yet another precursor of new physics, though I imagine Kelvin had in mind a mechanical effect of the ether on the matter, to be dealt with in the standard physics of the day. 

I put Milgrom's \cite{Milgrom} MOND (Modified Newtonian Dynamics) in this path. His idea that flat rotation curves of late-type galaxies could signify a departure from Newtonian gravity rather than the presence of dark matter seems quite out of line with the cosmological tests discussed in the last section, but MOND gives a remarkably successful prediction of the relation between circular velocity and detected baryonic mass in stars and neutral gas in these galaxies  \cite{McGaugh}. One has to be impressed that this prediction is observed over a far broader range of masses than was known when Milgrom proposed MOND.  We should support continued research on this cloud over  cosmology, by a few capable people, for  it  may teach us something of value. I offer in the next section two other anomalies that, whether real or apparent, also seem likely to be informative. 

\noindent 6. {\it Learning to compute.} Consider Kelvin's  opinion of heat capacities, and the unenthusiastic community response to MOND. We have a well-founded cosmology, just as Kelvin had a well-founded physics. Kelvin saw the possibility that computation of rates of relaxation to statistical equilibrium of internal energies of gasses might resolve the heat capacity puzzle. The cosmology community by and large is optimistic that better methods of computation will show that the standard cosmology predicts the observed baryon mass-circular velocity correlation, and the curious behavior of the dwarf galaxy spatial distribution and baryonic mass function \cite{dwarfs}, and other such issues. This echos the old admonition in physics, ``just trust the theory,'' which often proves to be a useful guide. Trust in the standard cosmology, as in explaining the MOND prediction within $\Lambda$CDM, is a good idea too, but for a different reason. The cosmological tests are far less dense than tests of the more well developed branches of natural science, and trust therefore more questionable. But by concentrating almost exclusively on the one standard cosmology rather than scattering attention across alternatives the community has a far better chance of discovering whether within the available evidence there really is something wrong with the chosen standard model. On the other hand, we should bear in mind the outcome of more flexible thinking about Kelvin's clouds. It is good strategy for some to scout more flexible thinking about cosmology, which may aid discovery of problems with given ideas, if there are any,  and advance discovery of remedies, if they are needed. 

\noindent 7. {\it Adding decimal places.} In Lord Kelvin's time some felt that all that was left to physics was adding decimal places. But improving accuracy with no other goal in mind is not to be deprecated: it can reveal anomalies. And despite the optimistic new title some give to our subject, precision cosmology, we still are proud to see measurements that get beyond the first decimal  place. 

\begin{figure}[h]
\begin{center}
\includegraphics[width=4in]{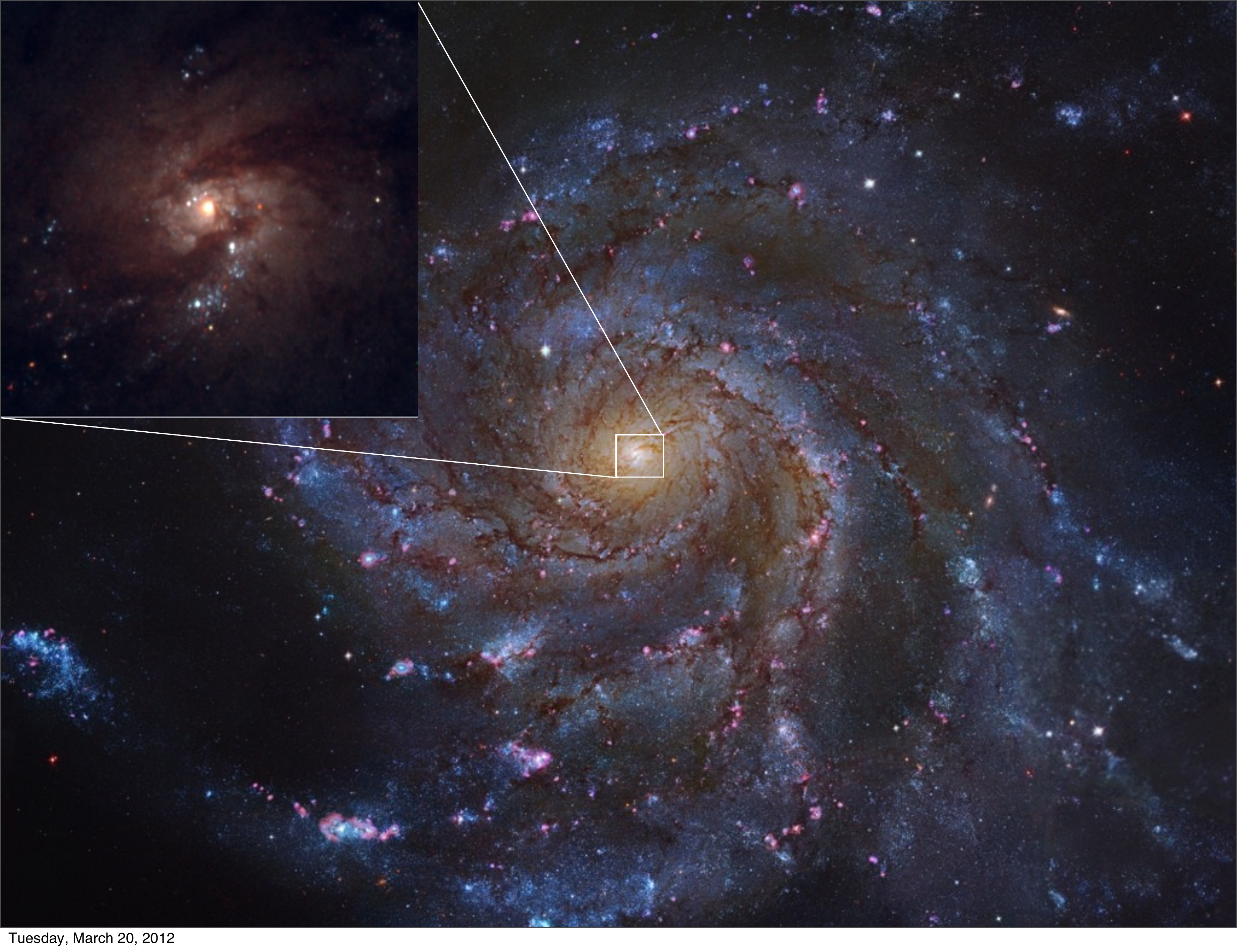}
\caption{The face-on pure disk galaxy M\,101. \label{Fig:M101}}
\end{center}
\end{figure}

\section{Island universes}

I offer two examples along path (5), apparent anomalies in the behavior of pure disk galaxies and elliptical galaxies. I mention these cases because both seem to me to be resistant enough to explanation within the  $\Lambda$CDM cosmology to be very likely to teach us something.

A pure disk galaxy is shown in  Figure~\ref{Fig:M101}. The main image of this nearby face-on spiral galaxy was made by R. Gendler by piecing together Hubble Space Telescope images and ground-based data. The dark lanes in the spiral arms are caused by obscuration by interstellar dust. The inset is a single HST image obtained by John Kormendy {\it et al.} \cite{Kormendy}. It shows that the dust lanes run inward to a central star cluster with radius $\sim 0.1$ percent of the radius of the luminous part of the galaxy. This galaxy appears to be in a near relaxed state, in which the dust will have settled onto the plane of the disk. The prominent dust lanes extending very close to the center of the galaxy indicate that the bulk of the stars also are confined to the plane, and that the structure of this galaxy is largely supported by rotation. Kormendy {\it et al.} \cite{Kormendy} argue that this situation does not seem to be uncommon:  they find that about half the 20 nearest big galaxies are largely supported by motions in the plane. 

The striking  whirlpool appearance of M\,101 seems to require that this galaxy grew by dissipative settling of gas or plasma onto the growing disk prior to significant star formation. If stars had formed in significant amounts in matter before it had settled into the disk it would have populated a stellar halo or classical bulge, which are not prominent components of this or other pure disk galaxies. 

Numerical simulations of structure formation in $\Lambda$CDM seem to predict that galaxies were assembled by hierarchical mergers of fragments (as illustrated in Fig.~3 in \cite{PeeblesNusser}). The larger fragments would have been massive and dense enough that pressure support required matter temperatures large enough to have collisionally ionized hydrogen. The plasma would lose energy by thermal bremsstrahlung, causing the fragments to contract until some energy source, presumably star formation, prevented it. But if too many stars formed in these fragments it would have produced a distinct stellar component outside the disk in the final product of the hierarchy of mergers. That is quite acceptable for some galaxies, such as our nearest large neighbor, M\,31, which has the substantial bulge of old stars that one might have expected to have resulted from  the merging seen in structure formation simulations. Recent simulations \cite{Brook,Guedes} do produce impressive pure disk galaxies, through careful attention to models for the conditions that promote and prevent star formation, but the mystery remains. The global star formation rate density was at its peak at redshifts $4\ga z\ga 2$. How could the progenitor fragments of pure disk galaxies have ``known'' not to have participated in this generally high global star formation rate?  One piece of the matter tumbling together according to the $\Lambda$CDM picture of the formation of the pure disk galaxy in Figure~\ref{Fig:M101}  ``knew'' it was going to host the growing disk, and start growing it at redshift well above unity if the age of the disk of the Milky Way \cite{AumerBinney} is typical of pure disk galaxies, while the rest of the fragments ``knew'' they had to hold off star formation until they had reached the growing disk. It's a curious situation.

\begin{figure}[h]
\begin{center}
\includegraphics[width=6.in]{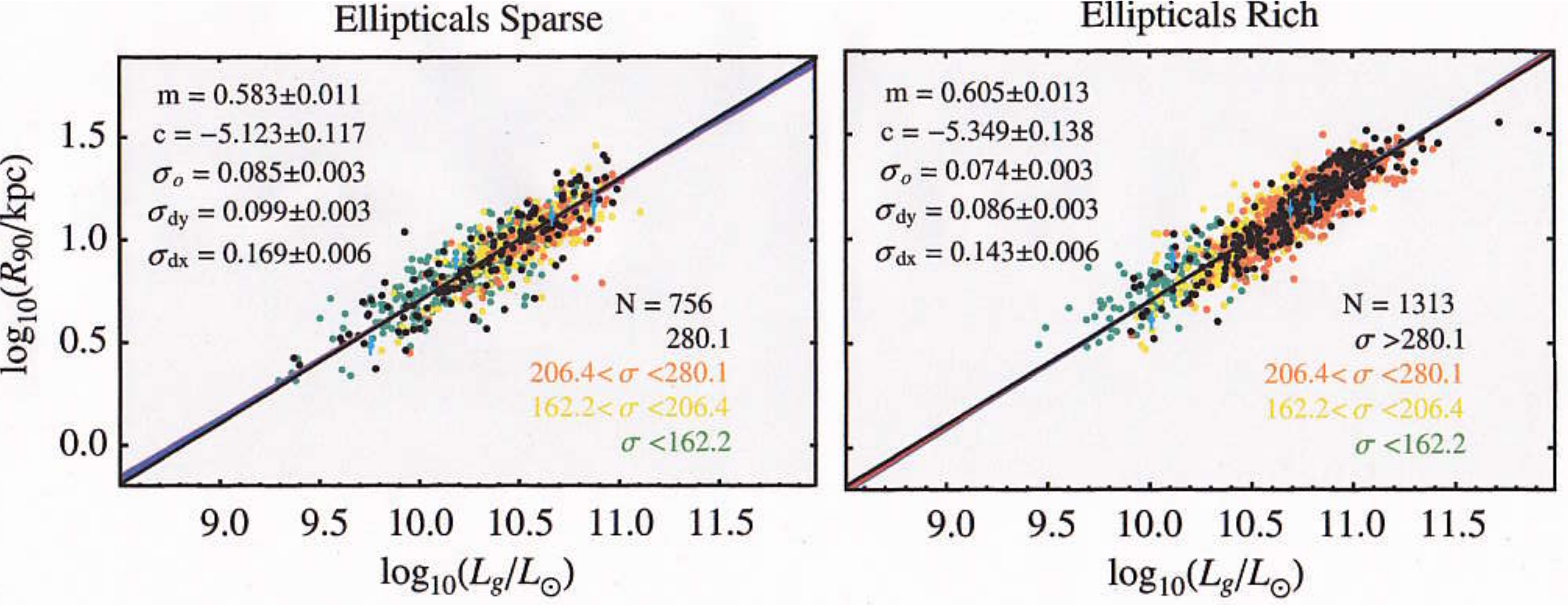}
\caption{Comparison of relations among elliptical galaxies found in regions of lower and higher  ambient densities.\label{Fig:Nair}}
\end{center}
\end{figure}

A possibly related issue is why overall properties of another large class of galaxies, ellipticals, are so little sensitive to environment. The phenomenon is illustrated in Figure~\ref{Fig:Nair}, taken from Nair, van den Bergh and Abraham \cite{Nair} (and based on the Sloan Digital Sky Survey). The vertical axis is a measure of radius. The authors took special care with this parameter because galaxies do not have edges: they just trail off into intergalactic space. The sample is separated according to ambient density, lower in the left-hand panel, higher in the right-hand panel. One sees several classical regularities. The Kormendy relation is the tendency for ellipticals with larger luminosities $L$ to have larger radii,  as indicated by the solid lines. The Faber-Jackson relation is the tendency for ellipticals with larger $L$ to have larger stellar velocity dispersions $\sigma$, as one can make out from the color coding for $\sigma$. The figure also illustrates the preference of more luminous  ellipticals for denser environments: in the higher ambient density sample in the right-hand panel the data points more densely populate the scatter plot near luminosity $\sim 10^{11}$ Solar luminosities and extend beyond that to luminosities three times the largest in the lower ambient density subsample. This preference  is predicted by the $\Lambda$CDM structure formation simulations. Also, Nair {\it et al.} \cite{Nair} find that radius $R_e$ that contains half the starlight of an elliptical shows more scatter, and more distinctly so in the lower density subset, than the radius $R_{90}$ used in Figure~\ref{Fig:Nair} that contains a larger fraction of the starlight. This is an environmental effect, perhaps rearrangement of internal structure by galaxy interactions \cite{Nair}. Environment certainly matters, but to me the striking and surely significant phenomenon is the overall insensitivity of the elliptical galaxy size-luminosity relation to the present environment.

\begin{figure}[h]
\begin{center}
\includegraphics[width=6in]{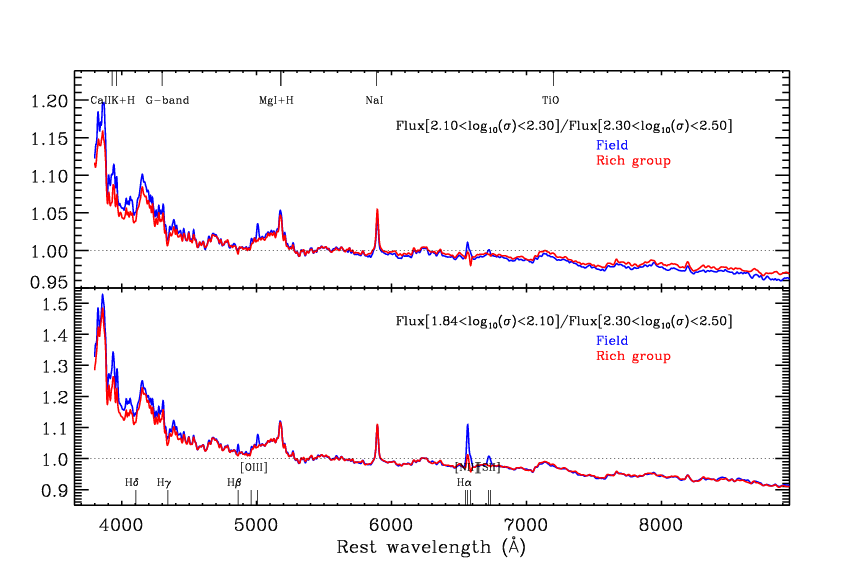}
\caption{Comparison of mean spectra of elliptical galaxies found in regions of lower and higher  ambient densities for three ranges of stellar velocity dispersion.\label{Fig:Gunagtun}}
\end{center}
\end{figure}

We see a similar phenomenon in the Zhu, Blanton and Moustakas \cite{Zhu} comparison of mean spectra of  central parts of ellipticals classified by the central stellar velocity dispersion $\sigma$ and the ambient density. Figure~\ref{Fig:Gunagtun} (kindly provided by Gunagtun Zhu using the results in \cite{Zhu}, and  based on SDSS) shows ratios of mean spectra in the three bins in $\sigma$ indicated in the two panels, separately for the higher and lower density subsamples. The ratios of spectra, with the lower $\sigma$  divided by the larger $\sigma$ mean spectrum, increase to the left, toward the blue. This illustrates the classical color-magnitude relation: more luminous ellipticals, that tend to have larger $\sigma$, tend to be redder. The variaton is larger in the lower panel that compares the lowest to highest $\sigma$ bins than in the upper panel that compares the intermediate to highest $\sigma$ bins: the relation is monotonic across the sample. The ratios for the lower density (field) and higher density (rich group) subsets are plotted in blue and red. The blue spike in the lower panel is the hydrogen recombination line H$\alpha$ likely from star-forming regions that are more common in ellipticals in lower density regions. The wonderful SDSS stability also reveals in these data a small but very clear relation between spectrum and ambient density, in the systematic tilt of blue relative to red curves. Figure~11 in \cite{Zhu} shows that at given $\sigma$ the spectra are bluer at lower density. There are exceptions to the very modest sensitivity to environment illustrated in Figure~\ref{Fig:Gunagtun}. The brightest galaxy in a cluster, which usually is near the cluster center defined by the X-ray emission \cite{BCG1}, tends to be redder than other early-type cluster members with the same $\sigma$ \cite{BCG2}, or it may host significant star formation \cite{Odea}. Both likely are  consequences of the special position at the bottom of a deep potential well and in a sea of hot plasma. That is, environment certainly matters to ellipticals. But again, to me the central point is that in the general class of ellipticals the variation of spectrum with environment is wonderfully small compared to the variation with velocity dispersion. 

We may say that Nair {\it et al.} classify ellipticals by a measure of internal conditions, the luminosity $L$, and a measure of external conditions, the ambient density. They find that another measure of internal conditions, the radius, is much more sensitive to $L$ than to the ambient density. Zhu {\it et al.} classify by the measure $\sigma$ of internal conditions and a measure of external conditions, and find that another measure of internal conditions, the spectrum, is  much more sensitive to $\sigma$ than ambient density. Other examples along the same lines are the relation between color and luminosity \cite{Hogg}, between red galaxy mass-to-light ratio and radius \cite{Bernardi,Magoulas},  and between galaxy stellar mass and radius \cite{Maltby}. One might read these results to mean that elliptical galaxy evolution has an attractor behavior that pulls ellipticals with given present $L$ to near common spectra, radii, and mass-to-light ratios. This is conceivable, and good science requires us to keep it in mind, but it seems unlikely to me. The alternative is that present environment does not much matter to ellipticals either because they formed at high redshift, when there was little variation in environment, and then passively evolved with little interaction with the environment, or else  they grew more slowly, but by rearrangement of matter that was already part of the protogalaxy evolving as a closed system. This straightforward reading of the evidence is not what pure dark matter simulations have led me to expect from $\Lambda$CDM.

Other broad classes of galaxies do show signatures of interaction with their environment. The transition of spiral galaxies into more nearly gas-free S0 galaxies seems to be a particularly clear example \cite{S0s, KormendyBender}. Galaxies are observed in the act of merging, and theory and intuition argue that the numerous close concentrations of galaxies at high redshift will have suffered many more mergers \cite{mergers}. These observations are in line with the hierarchical assembly of structures by mergers seen in simulations of the standard cosmology, and it is entirely reasonable therefore that mergers have figured heavily in discussions of how the galaxies formed.  Less reasonable, I think, is the scant attention to the evidence that two broad classes of galaxies, pure disks and ellipticals, appear to have evolved in near isolation from their surroundings, as island universes. This aspect of galaxy evolution also must be telling us something important about the origin of cosmic structure. 

\section{Concluding remarks}

Cosmology has advanced far more than I, or I suppose anyone, imagined a half century ago when I started looking into this subject. Perhaps witnessing the many wrong turns taken, and right turns overlooked so long, has conditioned me to be more doubtful about the stability of the present standard $\Lambda$CDM cosmology than I see in the thinking of many younger colleagues. But we can agree that the advances of the observational and theoretical bases are substantial and of lasting importance. This review leaves me with the impression that there is no shortage of interesting problems for study by the next generation, and I mention the island universe puzzle because it seems to me that the opportunities for research are even broader than generally advertised. 

\ack  I have benefitted from discussions with Michael Blanton, John Kormendy, Yen-Ting Lin, Piero Madau, Chris McKee, Surhud More, John Moustakas and Gunagtun Ben Zhu.

\end{document}